# Iterative Decoupling Method for High-Precision Imaging of Complex Surfaces

Eyal Baruch[1,a] and Izhak Bucher[1,b]

1 – Dynamics Laboratory, Faculty of Mechanical Engineering, Technion – Israel Institute of Technology, Haifa 3200003, Israel

[a]Electronic mail: eyalbaruch@campus.technion.ac.il

[b] Electronic mail: bucher@technion.ac.il

# Abstract

Nonlinear systems and interaction forces are pervasive in many scientific fields, such as nanoscale metrology and materials science, but their accurate identification is challenging due to their complex behaviour and inaccessibility of measured domains. This problem intensifies for continuous systems undergoing distributed, coupled interactions, such as in the case of topography measurement systems, measuring narrow and deep grooves. Presented is a method to invert a set of nonlinear coupled equations, which can be functions of unknown distributed physical quantities. The method employs a successive approach to iteratively converge to the exact solution of the set of nonlinear equations. The latter utilizes an approximate yet invertible model providing an inexact solution, which is evaluated using the hard-to-invert exact model of the system.

This method is applied to the problem of reconstructing the topography of surface contours using a thin and long vibrating fiber. In nanoscale metrology, measuring inaccessible deep and narrow grooves or steep walls becomes difficult and singular when attempting to extract distributed nonlinear interactions that depend on the topography. We verify our method numerically by simulating the Van der Waals (VdW) interaction forces between a nanofiber and a nanoscale deep groove, and experimentally by exploiting magnetic interactions between a magnetic topography and a vibrating, elastic beam. Our results validate the ability to accurately reconstruct the topography of normally inaccessible regions, making it a possible enhancement for traditional point based AFM measurements, as well as for other nonlinear inverse problems.

# 1. Introduction

A method for solving a set of coupled nonlinear equations is presented for physically inspired inverse problems. By exploiting a simplified inverse model, the method converges to the solution of the full model without explicitly inverting it. The method is implemented on a problem of reconstructing the topography of optically inaccessible topography domains. The suggested approach is inspired by Atomic Force Microscopes (AFMs) that employ a non-contact sensing by utilizing topography dependent potential forces and is thus targeted for nanoscale devices. As an example, sensing is accomplished by estimating the effect of distributed coupling forces on the natural frequency of a vibrating nanofiber. This method overcomes several obstacles, including smearing, time-consuming control methods, singular model inversion, and inaccurate inverse modeling.

System identification and parameter estimation are crucial and evolving areas of research [1], due to challenges regarding robustness to noise and singularity, and their imperative significance in the industrial sector for practical uses such as control and identification of physical features. A main challenge in this field is that the underlying equations can become singular when the measured data is not sufficiently informative. Additionally, even when the data is theoretically sufficient, in situations where the mapping between the measurements and fitted parameters is nonlinear [2–7], numerically inverting the system can become ill-posed or impossible. As a result, nonlinear system identification (e.g. [8] ) calls for a unique set of tools and is often restricted to a small number of parameters.

Numerous approaches have been proposed to address these challenges, ranging from direct, simple methods such as least squares [2], to complex, black-box methods such as deep neural networks. Reference [9] presented a method iteratively combining experimentation and post-processing to extract the dynamics of a given system. Reference [10] used genetic algorithms to determine the minimal set of dynamic equations that accurately reproduce measurements from a dynamic system. Reference [11] utilized a sparse set of modes to evolve the solution of partial differential equations over time, and [12] constructed a set of localized modes used to solve Schrödinger's equation for $N$ particles. In [13] it was proposed to build a library of functions for sparse identification of the model. Additionally, [14] developed an algorithm encompassing a wide range of methods, from dimensional analysis to neural networks, to enable system identification, while [15] integrated the partial differential equations of a dynamic system into the objective loss function of a neural network approximating the response of the dynamic system. Lastly, the process of modeling and identifying the dynamics of a system using a digital twin [16] was also explored.

Despite the mentioned apparent attempts, finding the inverse of a nonlinear multi-parameter problem remains a great challenge, especially where efficiency and speed are important. A central example for such systems is Atomic Force Microscopy (AFM) [18], which is a commonly used method for nanoscale metrology. This method utilizes Van der Waals (VdW) interactions between a sharp tip and the measured sample to reconstruct the topology of samples in atomic resolution. AFM is implemented using a variety of techniques, including Frequency Modulated AFM, which measures the change in the resonance frequencies of the system [19], due to changes in the overall potential affected by the spacing of interacting molecules in the sensor and specimen. However, AFM has several limitations. Firstly, the tip size makes it more difficult to measure complex, irregular topographies. This has been addressed partially by using different shaped tips [20–22] and tilting the AFM cantilever [23]. In spite of this, accurately measuring the contours of grooves remains a challenge. Secondly, AFM utilizes a point-by-point scan of the topography, tracking the natural frequency using phase-locked loop to find the distance of the cantilever from each point [24]. AFM based sensing is a time-consuming method, rendering it unusable for many applications. Thirdly, AFM works on the assumption that the tip interacts with a single point or atom per measurement [25]. It was shown that this assumption can be inaccurate, especially in groove-like topographies, resulting in significant smearing of the approximation in some cases [26].

This paper presents an approach to overcome some of these problems. By introducing a model-based successive approximation (MBSA) method, it is made possible to implicitly invert the set of coupled equations, using a simplified invertible model. For the AFM, this method is utilized to decouple distributed interactions between the topography and a thin, elongated vibrating fiber, which can be inserted into narrow and complex-shaped topographies. Successive approximation can be utilized to solve coupled nonlinear equations [27] and is implemented in a variety of applications, such as ADC converters [28], demosaicing [29], and optimal control design for nonlinear systems [30]. By decoupling the distributed interactions, the AFM method can accurately approximate grooves while reducing the smearing resulting from the localized interaction assumption. In addition, an approached based on introducing a model of the interactions is faster than the frequency tracking method, where the distance is measured by moving the sensing tip to a distance in which a fixed resonance frequency takes place. The process of moving the sensing tip with a control circuit is slow and it induces measurement uncertainties.

To verify the proposed method, a simulation of a resonating nanofiber measuring a groove was conducted. The distributed VdW interactions are successfully decoupled to accurately approximate the topography. In addition, an up-scaled experimental system that utilizes magnetic forces was constructed to realize the method experimentally. Since the magnetic potential is a function of the gap between material points on the nanofiber and the specimen to a certain negative power, similarly to the VdW potential, the approximation of a magnetic topography via magnetic interactions can be analogous to the exploitation of VdW interactions in AFM and is thus considered as a reasonable verification.

State of the art nanofiber based sensing and the prevalent FM-AFM [31] utilize perturbations of the system resonant frequency. Thus, Autoresonance (AR) was used to excite the system in resonance automatically at all configurations [32]. AR exploits a nonlinear element in the feedback loop (mathematically a signum function), and a 90 degrees phase shifting filter (normally an integrator), transiently inducing a positive feedback control loop to automatically excite a specific mode of vibration of the system in resonance [33,34]. The method does not require tuning of parameters like phase-locked loops and its convergence to the correct resonance frequency occurs immediately.

The paper is outlined as follows: Section 2 presents the MBSA method. Section 3 explores a case study of measuring the topography of a nanometric groove, modelling the interactions and implementing MBSA. Finally, Section 4 presents a numerical study where a simulated nanoscale groove topography is successfully reconstructed using MBSA. Finally, Section 5 presents experimental validation of the method using a magnetic beam and topography.

## 2. Model based successive approximation method

The proposed approach method of implicitly solving a set of nonlinear coupled equations is described. It is of interest to seek a set of discrete parameters denoted as $\mathbf{g}$ which reconstruct a distributed physical quantity (DPQ), denoted as $\mathbf{g}(\boldsymbol{\zeta})$. The property is distributed along a spatial coordinate vector $\boldsymbol{\zeta}$. The data exploited to obtain $\mathbf{g}$ is a set of measurements:

$$\boldsymbol{\omega}_d^2(\mathbf{x}) = \mathbf{f}(\mathbf{g}, \mathbf{x}), \qquad \boldsymbol{\omega}_d^2(\mathbf{x}) \in \mathbb{R}^m, \mathbf{g} \in \mathbb{R}^N \qquad (1)$$

where $\mathbf{f}$ is some known nonlinear model vector function of $\mathbf{g}, \mathbf{x}$ which cannot be explicitly inverted and $\mathbf{x}$ is a controllable parameter, e.g. the position of the sensor with respect to the sample. By incrementally changing $\mathbf{x}$, the vector $\boldsymbol{\omega}_d^2$ (the square of the measurement is used for convenience, as it describes the present case where the natural frequency squared is used for the approximation) is obtained.

Since an accurate representation of a DPQ requires a large number of parameters $g_i$, inverting the physical nonlinear model using standard methods such as gradient descent, can prove impossible due to the nonlinearity, complexity, and singularity of the inverted model. Singularity is manifested by large changes is the estimated parameters due small changes in the measured outcome, e.g. due to measurement noise. To overcome possible divergence due to singularity, an alternative approach is therefore presented.

Statement of the problem

Find $\mathbf{g}$ that minimizes: $\left\| \boldsymbol{\omega}_d^2 - \mathbf{f}(\mathbf{g}, \mathbf{x}) \right\|_2$

Solution strategy

A simplified implicitly invertible model $\mathbf{f_s}(\mathbf{g}, \mathbf{x})$ with identical inputs $\mathbf{g}$, and $\mathbf{x}$ is introduced. Since the model is inaccurate, the solution of the model with the correct $\mathbf{g}$ will yield:

$$\hat{\boldsymbol{\omega}}^2 = \mathbf{f}_s(\mathbf{g}, \mathbf{x}) \tag{2}$$

where the approximate model's outcome differs from the measured one, i.e. $\hat{\boldsymbol{\omega}}^2 \neq \boldsymbol{\omega}_d^2$ due to the simplification of the model. $\mathbf{f_s}(\mathbf{g}, \mathbf{x})$ can be inverted, producing an estimation of the topography, via:

$$\hat{\mathbf{g}} = \mathbf{f}_s^{-1}\left( \hat{\boldsymbol{\omega}}^2(\mathbf{x}), \mathbf{x} \right). \tag{3}$$

However, since $\mathbf{g}$ is unknown and generally $\hat{\mathbf{g}} \neq \mathbf{g}$, $\hat{\boldsymbol{\omega}}^2$ is also unknown. Thus, an iterative method is constructed to approximate $\hat{\boldsymbol{\omega}}^2$.

Iterations

For the $k_{th}$ iteration, given some $\hat{\boldsymbol{\omega}}_k^2$, an approximation of $\mathbf{g}$, denoted $\hat{\mathbf{g}}_k$ can be executed using Eq. (3):

$$\hat{\mathbf{g}}_k = \mathbf{f}_s^{-1}\left( \hat{\boldsymbol{\omega}}_k^2(\mathbf{x}), \mathbf{x} \right). \tag{4}$$

The following error function is devised:

$$\mathbf{e}_k = \boldsymbol{\omega}_d^2(\mathbf{x}) - \mathbf{f}\left( \hat{\mathbf{g}}_k, \mathbf{x} \right) \triangleq \boldsymbol{\omega}_d^2 - \hat{\boldsymbol{\omega}}_k^2 \tag{5}$$

where $\mathbf{f}$ is the full model and $\boldsymbol{\omega}_k^2$ is the approximation of the measurement vector assuming $\hat{\mathbf{g}}_k$ is exact. Assuming $\mathbf{f}$ is injective, $\mathbf{e}_k$ will converge to 0 only for $\hat{\mathbf{g}}_k = \mathbf{g}$. The error function is used to update $\hat{\boldsymbol{\omega}}^2$ in the following manner:

$$\hat{\boldsymbol{\omega}}_{k+1}^2 = \hat{\boldsymbol{\omega}}_k^2 + \beta \mathbf{e}_k. \tag{6}$$

Where $\beta$ is the parameter regulating the step size (see Appendix A). The algorithm is initialized using:

$$\hat{\boldsymbol{\omega}}_1^2 = \boldsymbol{\omega}_d^2 \,. \tag{7}$$

Equation (6) can be rearranged as:

$$\hat{\boldsymbol{\omega}}_{k+1}^2 = \hat{\boldsymbol{\omega}}_k^2 + \beta\left(\boldsymbol{\omega}_d^2 - \mathbf{f}\left(\mathbf{f}_s^{-1}\left(\hat{\boldsymbol{\omega}}_k^2, \mathbf{x}\right), \mathbf{x}\right)\right) \tag{8}$$

This method is applied iteratively until convergence. The condition for convergence (proof is presented in Appendix A) is:

$$\left(\frac{\partial \mathbf{f}_s}{\partial \mathbf{g}^T}\right)^T \frac{\partial \mathbf{f}}{\partial \mathbf{g}^T} \succ 0 \tag{9}$$

An example of the convergence method for a simple problem highlighting the various steps to convergence and its advantage on classical gradient descent based methods is presented in Appendix B. The approximation method is presented as a flowchart in Figure 1:

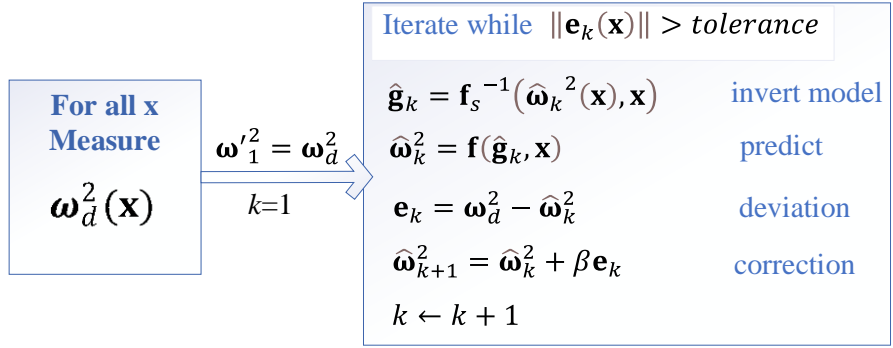

Figure 1 - A flowchart of the suggested successive approximation method

It is worth mentioning that the iterations are carried out in an offline manner and no repetitive experimentation is necessary beyond the one generating the measurement vector, $\boldsymbol{\omega}_d^2$. The latter introduces a great advantage over methods requiring additional experimentations [9] that are often slow and can introduce additional difficulties.

## 3. Nanofiber AFM topography reconstruction

The proposed method was applied on an AFM simulation of a nanofiber interacting with a groove topography. Therefore, a model of a system incorporating the VdW interaction forces between a resonating nanofiber and a specimen was developed. A schematic of the system and the sensing nanofiber is presented in Figure 2:

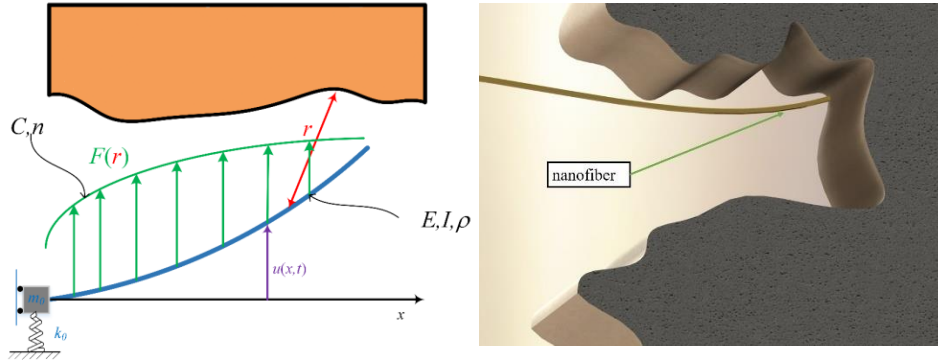

Figure 2 - The schematics of the simulated system. The oscillating fiber (in blue) interacts with the topography sample (in orange) resulting in distributed forces (in green) depending on the gap (r) between each point on the fiber and any point on the measured topography, illustrated by the red arrow. On the right, an artist's concept of the sensor and some illustrated topography is depicted.

The dynamics of this system are modeled in the following subsections.

### 3.1    Modeling interactions of a topography with a vibrating fiber.

Denoting $C$ and $n$ as the material constants describing the potential interaction for a specific configuration (see Figure 2), the potential of a 2D (the problem is constructed in 2D) topography element $ds$ acting on the point $x$ of the beam of the form [35,36]:

$$dV = -\frac{C}{r^n} ds \left[ kJ \right],$$  (10)

where $r$ represents the distance between a single point on the fiber, modeled as a Euler Bernoulli cantilever, and a single point on the measured sample (see Figure 2). The beam is parametrized as:

$$r_{beam} = \left( x, u_{ss}(x) + u(x,t) \right)$$  (11)

where $x$ is a coordinate along the fiber, $u_{ss}$ is the steady state static deflection of the beam caused by the static part of the interaction force, and $u$ is the beam's dynamic deflection coordinate caused by an oscillating excitation applied to it. The parametrization of the topography is:

$$r_{topography} = \left( g_1, g_2 \right),$$  (12)

and the surface contour of the 2D topography is parametrized as:

$$r_t = \left( g_1(\zeta), g_2(\zeta) \right),$$  (13)

where $\zeta \in [0, L_{top}]$ is a coordinate that follows the topography surface outline.

Substituting Eq. (11) and (12)s into (10) and integrating over the topography surface results in the potential element:

$$V = -\iint_{S} \frac{C}{\left(\sqrt{(g_1 - x)^2 + (g_2 - u - u_{ss})^2}\right)^n} dS \qquad (14)$$

Where $S$ is the topography surface interacting with the beam. Since the interacting forces are considerably smaller than the beam stiffness, it is assumed that $u_{ss} \approx 0$. Therefore, for small deviations from equilibrium, the second derivative of the potential gives us the added stiffness:

$$k(x) = -\iint_{S} \left( Cn \frac{(n+1) g_2{}^2 - (g_1 - x)^2}{\left(\sqrt{(g_1 - x)^2 + g_2{}^2}\right)^{n+4}} \right) dS \qquad (15)$$

It is assumed that $k(x)$, being relatively small compared to the bending stiffness, does not change the mode shapes significantly. Therefore, the displacement of a single mode can be approximated as:

$$u \approx a\phi(x). \qquad (16)$$

The modal equation of motion is:

$$\ddot{a}\phi + \omega_n^2 a\phi + \frac{k(x)}{\rho A} a\phi = 0 \qquad (17)$$

By multiplying by $\phi$ and integrating over the length of the beam, the following equation is obtained:

$$\ddot{a} + \left( \omega_n^2 + \frac{\int_0^{L_{beam}} \phi^2 k(x) dx}{\rho A \int_0^{L_{beam}} \phi^2 dx} \right) a = 0 \qquad (18)$$

As a result, the change in the natural frequency can be identified as:

$$\Delta \omega_n^2 = \frac{\int_0^{L_{beam}} \phi^2(x) k(x) dx}{\rho A \int_0^{L_{beam}} \phi^2(x) dx}. \qquad (19)$$

The latter is the Rayleigh quotient of the added stiffness due to the distributed interactions [38]. In addition to the fact that the mode shapes do not change significantly, their effect on the natural frequency exhibits low sensitivity to perturbations of the mode shape [40]. By measuring the natural frequencies of the system, information regarding $k(x)$ can be obtained. Subsequently, the topography contour $(g_1(\zeta), g_2(\zeta))$ can be approximated. However, this equation is a coupled nonlinear equation, since $\Delta\omega_{n,j}^2$ is a function of the distributed functions $(g_1, g_2)$ to a negative power. Using a standard gradient descent or Newton procedure to find $g(\zeta)$ is time consuming and leads to convergence and singularity problems (see Appendix B). The latter is because $g(\zeta)$ requires many parameters to accurately describe the topography. This necessitates the implementation of MBSA.

### 3.2 Successive topography approximation

To apply MBSA, the invertible model $\mathbf{f}_s(g(\zeta))$, is derived. This is done by assuming that firstly, $\left(g_1(\zeta), g_2(\zeta)\right)$ can be partitioned into sections which are presented either as $(g(\zeta), \zeta)$, where the tip moves perpendicular to the topography, or as $\left(\zeta, g(\zeta)\right)$, where the tip moves parallel to the topography. Secondly, it is assumed that $g(\zeta)$ can be approximated as a piecewise constant function:

$$g\left(\zeta\right) \approx \begin{bmatrix} g_1 & \cdots & g_i & \cdots & g_N \end{bmatrix}. \tag{20}$$

To further simplify the model, it is assumed that the strongest interaction between the topography and the beam is at $\zeta = 0$, where the denominator of the function under the integral is smallest. Therefore, other interactions are neglected, resulting in the invertible model:

$$\Delta\omega_{n,i}'^2 = -\frac{L_{top}}{N} Cn \frac{\int_0^{L_{beam}} \phi^2\left(x\right) \int_0^\infty \left(\dfrac{1}{\left(g_{1,i} + s_1 - x\right)^{n+2}}\right) ds_1 dx}{\rho A \int_0^{L_{beam}} \phi^2\left(x\right) dx} \tag{21}$$

where the integration to infinity is due to the assumption that $L_{top} \gg g_1$. Solving the integral results in:

$$\Delta\omega_{n,i}'^2 = -\frac{L_{top}}{N} \frac{Cn}{n+1} \frac{\int_0^{L_{beam}} \phi^2\left(x\right) \dfrac{1}{\left(g_{1,i} - x\right)^{n+1}} dx}{\rho A \int_0^{L_{beam}} \phi^2\left(x\right) dx} \tag{22}$$

Since the added stiffness becomes negligible far away from $x = L_{beam}$, $\phi^2(x)$ can be approximated as $\phi^2(x) \approx \frac{\phi^2(L_{beam})}{L_{beam}}$, resulting in:

$$\Delta\omega_{n,i}'^2 = -\frac{L_{top}}{N}\frac{C/(n+1)}{\bar{\bar{\phi}}\rho A\left(g_{1,i} - L_{beam}\right)^n}, \bar{\bar{\phi}} = \frac{1}{L_{beam}}\int_0^{L_{beam}}\left(\frac{\phi(x)}{\phi(L_{beam})}\right)^2 dx \tag{23}$$

Rearranging:

$$g_{1,i} = L_{beam} + \left(-\frac{L_{top}}{N}\frac{C/(n+1)}{\bar{\bar{\phi}}\rho A\Delta\omega_{n,i}'^2}\right)^{1/n} \tag{24}$$

Similarly, if the fiber is parallel to the topography, taking the topography closest to the tip ($g_1 = L_{beam}$) and assuming it interacts solely with the tip results in the stiffness:

$$k(x) = \int_0^\infty \left(-\frac{Cn(n+1)}{\left(g_2 + s_2\right)^{n+2}}\right)ds_1\delta\left(x - L_{beam}\right) \tag{25}$$

where the integration to infinity is due to the assumption that $L_{top} \gg g_2$. In addition $\delta(x - L_{beam})$ is the Dirac delta function, which is as a result of taking into account solely the interaction with the tip. Substituting in Eq. (19) and rearranging:

$$g_2 = \left(-\frac{CL_{top}}{N\bar{\bar{\phi}}\Delta\omega_{n,i}'^2}\right)^{1/n}. \tag{26}$$

Since the main interaction forces are resulting from the portion of the topography which is parallel to fiber, in can be induced that the interacting force is proportional to the depth in which the beam is in the groove. Therefore Eq. (26) can be modified to take the depth into account:

$$g_2 = \left(-\frac{CL_{top}d}{N\bar{\bar{\phi}}\Delta\omega_{n,i}'^2}\right)^{1/n} \tag{27}$$

where $d$ is the depth in the groove. Note that the invertible system is a set of decoupled nonlinear equations of the single parameter $g_i$ which can be easily obtained. By measuring the change in the natural frequency corresponding to each piecewise constant segment $g_i$ in the topography, the set of equations (26) and (27) can be solved, resulting in an approximation of the topography **g′**. The measured perturbation in the natural frequency is used for the iterative approximation method, where **g′** converges to the real topography **g**. It is important to note that for the assumption of Eq. (16) to be valid, AR was used to measure the natural frequency of the beam. AR allows the automatic locking onto the system's resonance frequency [32,34,41]. By taking the response of the system, phase shifting it in 90 degrees and putting it through a relay (a signum function), the system is automatically excited in resonance.

## 4. Simulation results

A simulation of a gold nanofiber interacting with a silicon made topography, was conducted. the constants and parameters of the simulation and their derivation are detailed in Appendix C. The approximation of the topography was executed in the following steps: first, the natural frequency of the fiber, when vibrating perpendicular to the topography, was measured (see Figure 3.A, and Video 1). Next, the measurement of the natural frequency at each point is used to approximate the perpendicular part of the topography and indicate the existence of a groove, as presented in Figure 3.B (see Video 2), using MBSA. Then, the lower sidewall was measured via the change in the natural frequency of the fiber (see Figure 4.A and Video 3), and the sidewall was also approximated (see Figure 4.B and Video 4). In the same manner, the upper sidewall and the base were approximated, resulting in the full approximation of the topography (Figure 5).

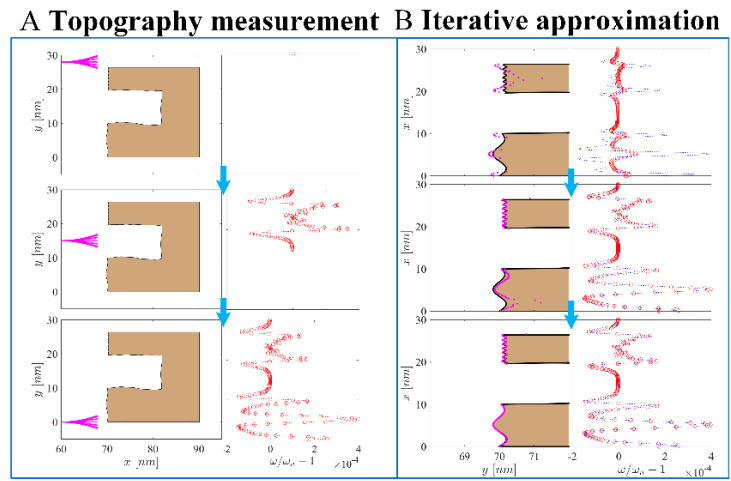

Figure 3 - The approximation method. Section A shows the measurement of the natural frequency (in red) of the fiber (in magenta) when positioned perpendicular to different points in the topography (in brown). Section B shows the iterative convergence of the topography approximation (in magenta) to the exact topography (in brown) simultaneously with the convergence of the approximated frequency vector (in blue) to the measured one (in red).

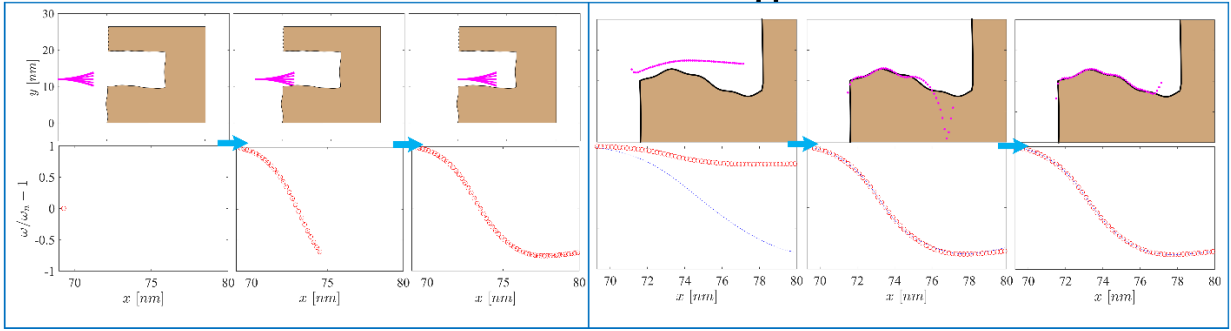

Figure 4 – The sidewall approximation. Section A shows the measurement of the natural frequency (in red) of the fiber (in magenta) when positioned parallel to different points in the lower sidewall of the topography (in brown). Section B shows the convergence of the approximation of the sidewall to the exact topography.

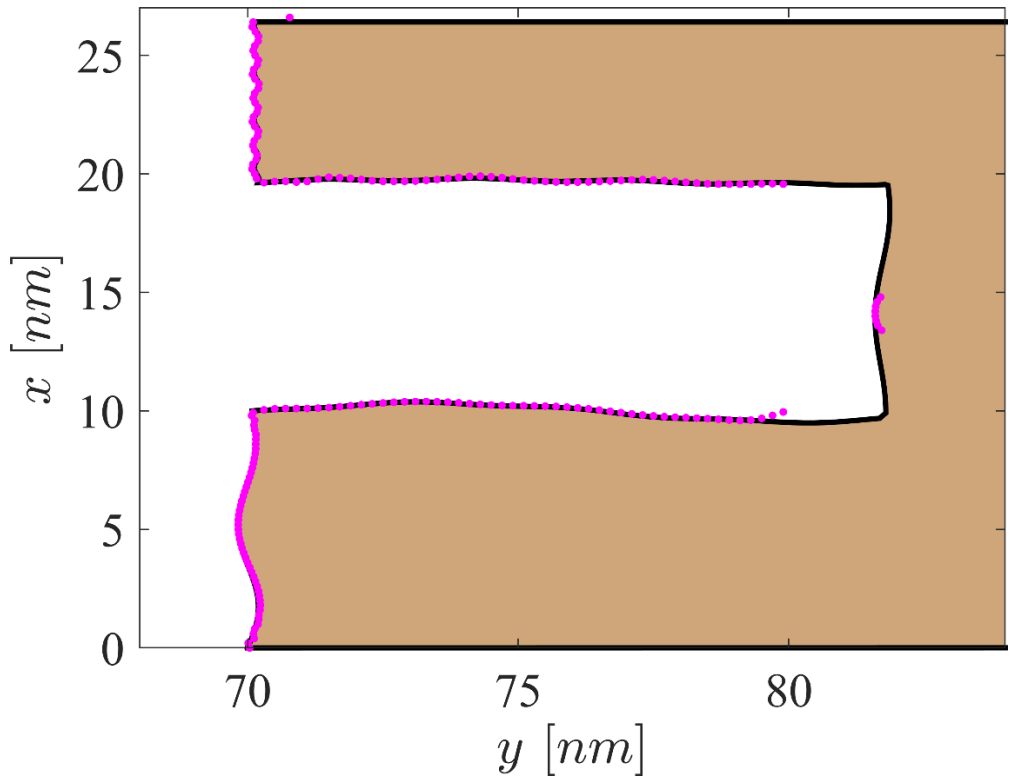

Figure 5 - the full topography approximation.

## 5. Experimental verification

The proposed method was verified using a magnetic system (see Figure 6) which substitutes the VdW potential with a magnetic one since they have similar characteristics in certain configurations.

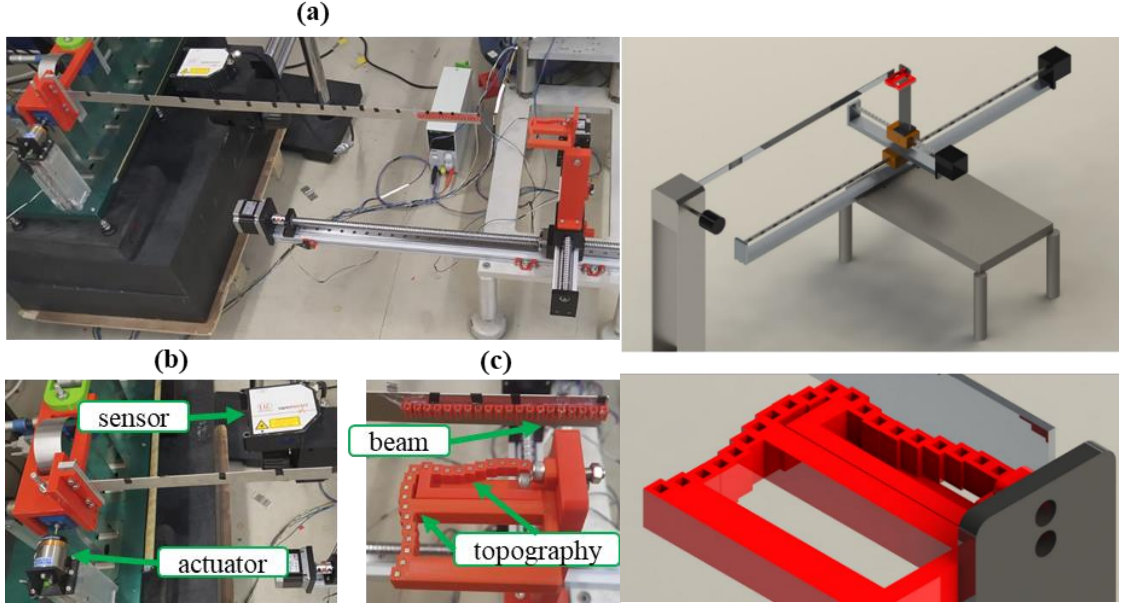

Figure 6 - The magnetic experimental system. (a) The full system. (b) The sensor measuring the beam vibration and the actuator applying the sinusoidal excitation. (c) The measured topography and the magnetic beam. The potential constants were curve-fitted, and the topography was approximated via the calibrated potential

The actuated beam that substitutes the AFM nanofiber is a 0.682 meters long aluminum beam, with a rectangular cross-section of 21 by 1 mm. An array of 11 magnets spaced in intervals of 5 mm is attached to the beam. the measured magnetic topography, is 2 connected arrays of magnets, one perpendicular to the beam and the other parallel. The magnets are also spaced in intervals of 5 mm with a sinusoidal contour with the amplitude of 2.5 mm. The polarity of the beam magnets is opposite to the one of the topography magnets. Since the interactions are discrete, and are a result of the topography surface, Eq. (15) becomes:

$$k\left(x_j\right) = \sum_{i=1}^{N} \left( Cn \frac{\left(g_1\left(\zeta_i\right) - x_j\right)^2 - (n+1)\left(g_2\left(\zeta_i\right) - u_{ss}\right)^2}{\left(\sqrt{\left(g_1\left(\zeta_i\right) - x_j\right)^2 + \left(g_2\left(\zeta_i\right) - u_{ss}\right)^2}\right)^{n+4}} \right) \tag{28}$$

Where $i$ represents the $i_{th}$ magnet of the topography, and $j$ represents the $j_{th}$ magnet of the beam. This formulation holds under the assumption that Eq. (10) holds. To justify this assumption and evaluate the values of $C$ and $n$, a model of the magnetic potential was devised.

### Magnetic potential

A model of the magnetic forces in the experimental system was derived. The interaction potential between two dipoles can be approximated as [44]:

$$V = -\frac{C}{r^3}\left(3\left(I_1 \cdot \hat{r}\right)\left(I_2 \cdot \hat{r}\right) - \left(I_1 \cdot I_2\right)\right)$$

$$(29)$$

where $C$ is a constant resulting from the magnetic moments and permeability, $r$ is the distance between the two centers of the moments, $\hat{r}$ is the normalized vector in the direction of $r$, and $I_1, I_2$ are the normalized vectors in the directions of the first and second dipoles respectively. Since the magnets can be approximated as dipoles, it is assumed that the vectors $I_1, I_2$ between a topography magnet and a beam magnet are parallel with opposite signs, and that $\hat{r}$ is orthogonal to $I_1, I_2$, the force becomes:

$$V = -\frac{C}{r^3}.$$

$$(30)$$

### Magnetic potential calibration

For the model to better suit the experimental system, the constants $C$ and $n$ (the power of $r$) were calibrated using a single magnet. The natural frequency of the beam was measured at different known distances from the beam. Eq. (19) was used to approximate and the values of $C$, $n$ and $\omega_0$ using nonlinear least squares, with **g** being a known topography. The calibration set and its approximation are presented in Figure 7:

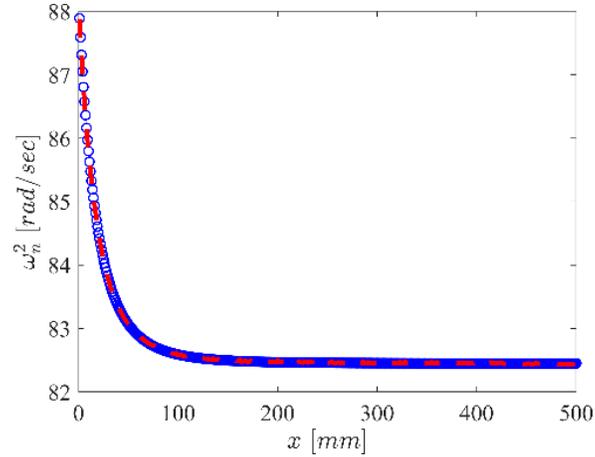

Figure 7– The calibration set. In red is the measured set. In blue is the approximated set using the calibrated parameters.

The error between the measured and approximated natural frequency is small, indicating a precise approximation of $C$ and $n$. The values extracted from the approximation are:

$$\begin{aligned} C &= 67981 \\ n &= 3.356380 \end{aligned} \tag{31}$$

The power is relatively close to the one derived analytically (see Eq. (30)), validating the potential model.

### 5.3 Experimental results

The approximation set was measured by moving the beam across the topography at a nominal distance of 20 [mm] and measuring the natural frequency (see Video 5). Using the measurements, the perpendicular topography was approximated using MBSA methodology (see Figure 8 and Video 6), and the beam was inserted parallel to the topography, to approximate the groove sidewall (see Figure 9 and Video 6). The full topography is constructed from the two approximated parts, as presented in Figure 10:

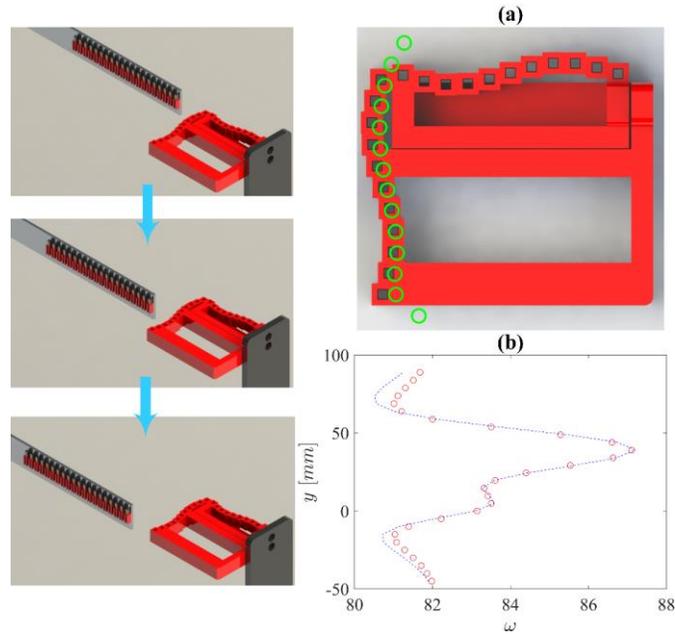

Figure 8 - The topography approximation process. The left side depicts the measurement process of the perpendicular part of the topography. (a) shows the approximation results, and (b) shows the approximated and measured frequency vector

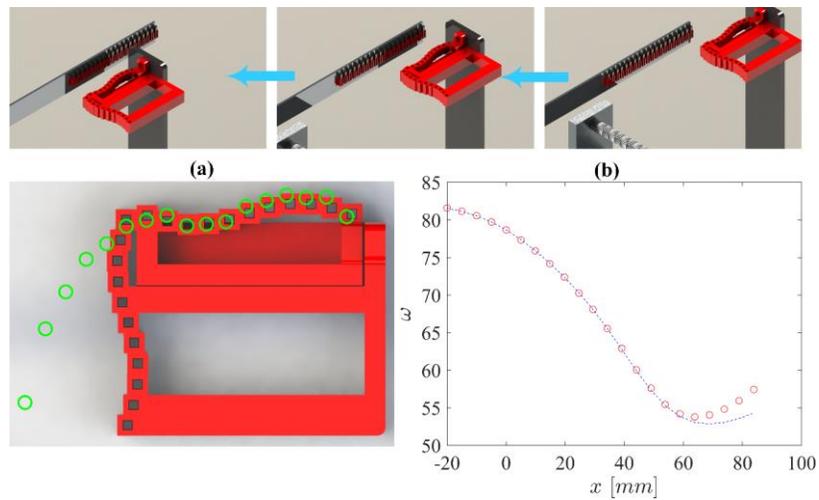

Figure 9 - The topography approximation process. The upper panels depict the measurement process of the parallel part of the topography. (a) shows the approximation results, and (b) shows the approximated and measured frequency vector

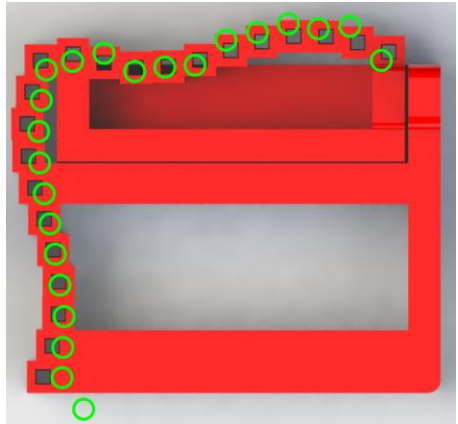

Figure 10 - The full topography approximation, using MBSA.

The error percentages of the approximation are presented in Figure 11:

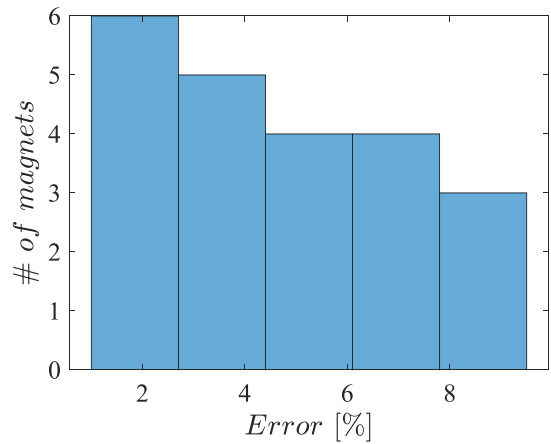

Figure 11 – A histogram of the errors for each magnet.

It is evident that MBSA approximates the measured topography with reasonable accuracy, overcoming the coupled interactions. Residual magnetic fields and geometrical errors may cause some deviations from the true magnets and their estimated locations. Additionally, the variance in the magnetic potential of the individual magnets adds to the inaccuracy. It is expected that nanoscale, the atomic or molecular interactions can be better described with the proposed approach, leading to a more accurate approximation.

## 6. Discussion and conclusions

A method for inverting a set of nonlinear equations and thus accurately approximating a measured topography from coupled interaction forces with a resonating fiber was presented. The method makes use of an accurate model of the interactions, which was derived as the solution to the Rayleigh quotient of a beam with a distributed added stiffness. To do so, a simplified model which takes into account a single interaction and is therefore invertible, was developed. The method minimizes the error between the measured natural frequency of the beam and the approximation of the same vector derived by applying the accurate model on the topography approximation of the simplified one. The criterion that guarantees the convergence of the method was presented, showing that the convergence is dependent on the partial derivatives of the simplified model the and accurate one.

The procedure was validated on a simulation of a thin gold nanofiber interacting with a silicon groove. The ability to accurately approximate both the outer surface contour and the groove sidewalls waws demonstrated. It is assumed that taking multiple walls into account in the full model will enable the accurate approximation of the groove base and its edges as well.

In addition, MBSA was implemented on an up-scaled system that substitutes VdW interaction forces with magnetic ones. The potential magnetic constants were calibrated by measuring a known topography and calibrating the constants of the model to match the results. Then, the beam was autoresonated to measure its natural frequency. The results display the ability to accurately decouple distributed interactions and solve a nonlinear set of equations. The advantage of MBSA is made clear by the contrast between the initial approximation, which could be the result of standard AFM measurements, and the final one after optimization.

# Appendix A – An iterative approximation method for the solution of a set of nonlinear equations

It is of interest to solve the set of nonlinear equations, which can be represented as:

$$\mathbf{f}\left(\mathbf{g}, \mathbf{x}\right) = \boldsymbol{\omega}_d^2, \quad \boldsymbol{\omega}_d^2 \in \mathbb{R}^N, \quad \mathbf{g} \in \mathbb{R}^M \tag{32}$$

Where $\mathbf{f}$ represents the system model, which in the context of this paper is represented in Eq.(13), $\mathbf{g}$ is the unknown piecewise constant parametric topography vector we wish to approximate, and $\boldsymbol{\omega}_d^2$ is the measurement vector used to approximate $\mathbf{g}$. It is assumed that $\mathbf{f}$ is an injective function so that every value of $\boldsymbol{\omega}_d^2$ corresponds to a single $\mathbf{g}$.

While $\mathbf{f}$ is not easily invertible, there is a different equation:

$$\mathbf{f}_s\left(\mathbf{g}, \mathbf{x}\right) = \hat{\boldsymbol{\omega}}^2 \tag{33}$$

which is solvable for $\mathbf{g}$. However, the vector $\hat{\boldsymbol{\omega}}^2$ is unknown. Therefore, it is of interest to find the vector $\hat{\boldsymbol{\omega}}^2$ for which Eq. (33) will result in the correct $\mathbf{g}$. Having found $\hat{\boldsymbol{\omega}}^2$, $\mathbf{f}_s$ can be inverted in a straight forward and numerically robust manner to find an approximation of $\mathbf{g}$.

This will be done using 4 iterative steps:

Step 1: Solve Eq. (33) for $\mathbf{g} = \hat{\mathbf{g}}_k$ using the estimate of $\hat{\boldsymbol{\omega}}^2$ at the $k^{\text{th}}$ iteration, denoted, $\hat{\boldsymbol{\omega}}_k^2$, i.e:

$$\hat{\mathbf{g}}_k = \mathbf{f}_s^{-1}\left(\hat{\boldsymbol{\omega}}_k^2, \mathbf{x}\right) \tag{34}$$

Step 2: Calculate the deviation in $\boldsymbol{\omega}$ from the true value $\boldsymbol{\omega}_d^2$ (note that $\mathbf{f}(\hat{\mathbf{g}}_k, x)$ is exact and can be evaluated directly):

$$\mathbf{e}_k = \boldsymbol{\omega}_d^2 - \mathbf{f}\left(\hat{\mathbf{g}}_k, \mathbf{x}\right) \tag{35}$$

Step 3: Update the input function to Eq. (34):

$$\hat{\boldsymbol{\omega}}_{k+1}^2 = \hat{\boldsymbol{\omega}}_k^2 + \Delta\hat{\boldsymbol{\omega}}^2 \tag{36}$$

It should be stated that $\Delta\hat{\boldsymbol{\omega}}^2$ is some a function of the error, which will be chosen so that the norm of the error $J = \left\|\mathbf{e}\right\|_2$ decreases.

Step 4: Repeat. Steps 1-3 using $\hat{\boldsymbol{\omega}}_{k+1}{}^2$, until $J < \varepsilon_e$.

### A.1     Conditions for convergence.

Taking the $k^{\text{th}}$ iteration, where the first iteration is:

$$\hat{\boldsymbol{\omega}}_1^2 \overset{\Delta}{=} \boldsymbol{\omega}_d^2 \tag{37}$$

Eq. (33) can be solved, resulting in:

$$\hat{\mathbf{g}}_k = \mathbf{f}_s^{-1}\left(\hat{\boldsymbol{\omega}}_k^2\right) \tag{38}$$

The calculated $\hat{\mathbf{g}}_k$ can be substituted into Eq.(32), resulting in the error of the full equation (Eq. (32)):

$$\mathbf{e}_k = \boldsymbol{\omega}_d^2 - \mathbf{f}\left(\hat{\mathbf{g}}_k, \mathbf{x}\right) \tag{39}$$

The error of the solvable equation (Eq. (33)) is:

$$\mathbf{e}_{s,k} = \hat{\boldsymbol{\omega}}^2 - \hat{\boldsymbol{\omega}}_k^2 \tag{40}$$

The system will converge if

$$\left\|\mathbf{e}_{s,k+1}\right\|_2 < \left\|\mathbf{e}_{s,k}\right\|_2 \tag{41}$$

Substituting Eq. (40) into (41):

$$\left\|\mathbf{e}_{s,k+1}\right\|_2^2 = \left(\hat{\boldsymbol{\omega}}^2 - \hat{\boldsymbol{\omega}}_k^2 - \Delta\hat{\boldsymbol{\omega}}^2\right)^T \left(\hat{\boldsymbol{\omega}}^2 - \hat{\boldsymbol{\omega}}_k^2 - \Delta\hat{\boldsymbol{\omega}}^2\right) = \left\|\mathbf{e}_{s,k}\right\|_2^2 + \left\|\Delta\hat{\boldsymbol{\omega}}^2\right\|_2^2 - 2\mathbf{e}_{s,k}^T\Delta\hat{\boldsymbol{\omega}}^2 \tag{42}$$

For Eq. (41) to hold, the following must be dictated:

$$\left\|\Delta\hat{\boldsymbol{\omega}}^2\right\|_2^2 < 2\mathbf{e}_{s,k}^T\Delta\hat{\boldsymbol{\omega}}^2 \tag{43}$$

By choosing:

$$\Delta\hat{\boldsymbol{\omega}}^2 = \beta\mathbf{e}_k, \beta > 0 \tag{44}$$

And substituting into Eq. (43):

$$\beta \left\| \mathbf{e}_k \right\|_2^2 < 2\mathbf{e}_{s,k}^T \mathbf{e}_k \tag{45}$$

From Eq. (45) it can be deduced that:

$$\beta < \left| \frac{2\mathbf{e}_{s,k}^T \mathbf{e}_k}{\mathbf{e}_k^{\ T} \mathbf{e}_k} \right| \approx 2 \tag{46}$$

In practice, $\beta$ was chosen to be a smaller value to guarantee convergence. Since $\beta$ can be chosen to be arbitrarily small, it is enough to show that

$$0 < 2\mathbf{e}_{s,k}^T \mathbf{e}_k. \tag{47}$$

This choice of $\Delta \hat{\boldsymbol{\omega}}^2$ allows us to rearrange Eq. (36) as:

$$\hat{\boldsymbol{\omega}}_{k+1}^2 = \hat{\boldsymbol{\omega}}_k^2 + \beta \left( \boldsymbol{\omega}_d^2 - \mathbf{f} \left( \mathbf{f}_s^{-1} \left( \hat{\boldsymbol{\omega}}_k^2, \mathbf{x} \right) \right) \right) \tag{48}$$

Using the fundamental theorem of calculus:

$$\mathbf{e}_k = \mathbf{f}(\mathbf{g}) - \mathbf{f}(\hat{\mathbf{g}}_k) = (\mathbf{g} - \hat{\mathbf{g}}_k) \int_0^1 \frac{\partial \mathbf{f}}{\partial \mathbf{g}} \left( \mathbf{g} + \lambda (\mathbf{g} - \hat{\mathbf{g}}_k) \right) d\lambda, \ \lambda \in \left[ \hat{\mathbf{g}}_k, \mathbf{g} \right]$$

$$\mathbf{e}_{s,k} = \mathbf{f}_s(\mathbf{g}) - \mathbf{f}_s(\hat{\mathbf{g}}_k) = (\mathbf{g} - \hat{\mathbf{g}}_k) \int_0^1 \frac{\partial \mathbf{f}_s}{\partial \mathbf{g}} \left( \mathbf{g} + \lambda (\mathbf{g} - \hat{\mathbf{g}}_k) \right) d\lambda, \ \lambda \in \left[ \hat{\mathbf{g}}_k, \mathbf{g} \right] \tag{49}$$

Eq. (47) is therefore:

$$(\mathbf{g} - \hat{\mathbf{g}}_k)^T \int_0^1 \left[ \frac{\partial \mathbf{f}_s}{\partial \mathbf{g}^T} \left( \mathbf{g} + \lambda (\mathbf{g} - \hat{\mathbf{g}}_k) \right) \right] d\lambda \int_0^1 \left[ \frac{\partial \mathbf{f}}{\partial \mathbf{g}^T} \left( \mathbf{g} + \lambda (\mathbf{g} - \hat{\mathbf{g}}_k) \right) \right] d\lambda (\mathbf{g} - \hat{\mathbf{g}}_k) > 0 \tag{50}$$

using Taylor expansion, assuming $\mathbf{g} - \mathbf{g}_k$ is relatively small:

$$\mathbf{e}_k = \mathbf{f}(\mathbf{g}) - \mathbf{f}(\hat{\mathbf{g}}_k) \approx \frac{\partial \mathbf{f}}{\partial \mathbf{g}} (\mathbf{g} - \hat{\mathbf{g}}_k)$$

$$\mathbf{e}_{s,k} = \mathbf{f}_s(\mathbf{g}) - \mathbf{f}_s(\hat{\mathbf{g}}_k) \approx \frac{\partial \mathbf{f}_s}{\partial \mathbf{g}} (\mathbf{g} - \hat{\mathbf{g}}_k) \tag{51}$$

Eq. (47) can be approximated as:

$$(\mathbf{g} - \hat{\mathbf{g}}_k)^T \left( \frac{\partial \mathbf{f}_s}{\partial \mathbf{g}} \right)^T \frac{\partial \mathbf{f}}{\partial \mathbf{g}} (\mathbf{g} - \hat{\mathbf{g}}_k) > 0 \tag{52}$$

Therefore, Eq. (52) holds iff the matrix $\left( \dfrac{\partial \mathbf{f}_s}{\partial \mathbf{g}^T} \right)^T \dfrac{\partial \mathbf{f}}{\partial \mathbf{g}^T}$ is positive definite for all $\mathbf{g}$. In addition, when the system converges:

$$\hat{\boldsymbol{\omega}}_{k+1}^2 = \hat{\boldsymbol{\omega}}_k^2 \rightarrow \Delta\hat{\boldsymbol{\omega}}^2 = 0 \,. \tag{53}$$

This happens only when $\mathbf{e}_i = 0$, which results in

$$\hat{\mathbf{g}}_k = \mathbf{g} \tag{54}$$

since the system is injective.

Therefore, if Eq. (47) holds, the method will converge to the desired $\mathbf{g}$.

## Appendix B – An illustrative example of the implementation of MBSA

Consider the single parameter function:

$$f\left( x \right) = \sin(15x) + 8x + 3 \tag{55}$$

It is of interest to approximate the solution of the problem:

$$f\left( x \right) = 0 \tag{56}$$

By implementing gradient descent, with the naïve initial guess of $x_0 = 0$, the optimization method converges to a local minimum. i.e $(0.1409, 1.1065)$ (see Figure 12.a). This is a sub-optimal result. However, by implementing a simplified invertible model:

$$f_s\left( x \right) = 6x = \omega, \quad \omega_0 = 0 \tag{57}$$

For which the inverse is:

$$f_s^{-1}\left( \omega \right) = \frac{\omega}{6} = x_s \tag{58}$$

it is possible to implement MBSA. Notice that the inverse model is inexact, neglecting the sinusoidal component and DC component and inaccurately approximating the gradient of the linear component. However, it still maintains the general gradient of the function.

Following Eq. (48), the iterative law will be:

$$\omega_{k+1} = \omega_k + \beta\left(\omega_n - f\left(\frac{\omega_k}{6}\right)\right) = \omega_k + \beta\left(\omega_n - \left(\sin\left(\frac{15\omega_k}{6}\right) + \frac{8\omega_k}{6} + 3\right)\right)$$

(59)

This method results in the convergence to the desired solution, as presented in Figure 12.b

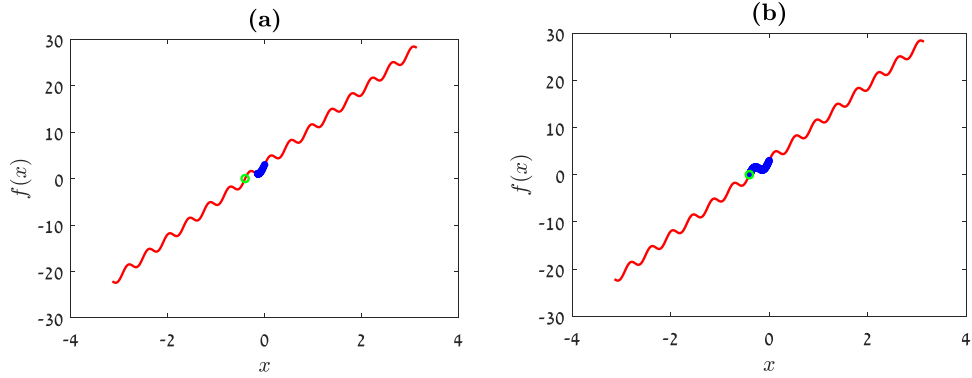

Figure 12 - Convergence of the different optimization methods searching for a result to the equation $sin(5x) + 8x + 3 = 0$. The optimized function is in red, the optimization steps are in blue, and the optimal point (0) is in green. (a): Gradient based optimization. (b): MBSA optimization

To prove the convergence of MBSA in this problem, it will be shown that Eq. (47) holds. Substituting Eq. (39) and (40) into Eq. (47):

$$\mathbf{e}^T\mathbf{e}_s = \left(8(x - x_d) + 2\cos\left(\frac{15}{2}(x + x_d)\right)\sin\left(\frac{15}{2}(x - x_d)\right)\right)6(x - x_d)$$

(60)

For $|x - x_d| > 1/4$ we get

$$2\beta\mathbf{e}^T\mathbf{e}_s > \varepsilon\left(x - x_d\right)^2, \quad \varepsilon > 0$$

.

(61)

Which results in the system converging. However, for

$$\left(8(x - x_d) + 2\cos\left(\frac{15}{2}(x + x_d)\right)\sin\left(\frac{15}{2}(x - x_d)\right)\right)6(x - x_d) < 0$$

(62)

And $|x - x_d| < 1/4$, the system diverges. Now, applying Taylor expansion in the form of:

$$\sin\left(\frac{15}{2}(x - x_d)\right) \approx \frac{15}{2}(x - x_d) \triangleq \frac{15}{2}\Delta x$$

(63)

Results in

$$\left(8 + 15\cos\left(15\left(x_d + \Delta x\right)\right)\right)6\Delta x^2 < 0 \tag{64}$$

Note that the function in brackets is $f'(x)$, which results in the state: $f(x_d) = 0$, $f'(x_d) < 0$. However, since it is known that there exists some $x$ for which $f(x) > 0$, there must be another crossing of $f(x_{d,2}) = 0$, $f'(x_{d,2}) > 0$, for which Eq. (47) does hold, resulting in the convergence to $x_{d,2}$. As a result, the system converges to the point $f(x_d) = 0$ for any initial condition.

It is worth noting that this result holds regardless of the amplitudes of the sinusoidal and linear components. In addition, while MBSA was demonstrated on a single parameter system, this method can be extended to multiple parameters, provided an invertible model with similar gradients is constructed.

## Appendix C - Method simulation constants

For the simulation, the interaction forces between a gold nanowire and a silicon topography were implemented using the Lennard – Jones approximation of the Van der Waals Potential [35,36]:

$$V = -4\varepsilon\left(\frac{\sigma}{r}\right)^6 \left[kJmol^{-1}\right] \tag{65}$$

The values of the parameters $\varepsilon$ and $\sigma$ for gold and silicon are [42]:

| Constants | Gold | Silicon |
|---|---|---|
| $\sigma[nm]$ | 0.293373 | 0.392 |
| $\varepsilon\left[\dfrac{kJ}{mol}\right]$ | 0.163176 | 2.51040 |

The constants $\sigma_{s-g}, \varepsilon_{s-g}$ which are used to construct the potential acting between silicon and gold, can be approximated as [43]:

$$\sigma_{s-g} = \frac{1}{2}\left(\sigma_s + \sigma_g\right) \tag{66}$$

$$\varepsilon_{s-g} = \sqrt{\varepsilon_s \varepsilon_g} \tag{67}$$

## Acknowledgments

The Authors would like to thank the valuable comments of Professor Keegan Moore and the preparatory work done by Mr. Nir Ben Shaya in his MSc research, supporting some of the initial work presented here.